\documentstyle[sprocl]{article}

\newcommand{\msb}{\mbox{$\overline{\rm{MS}}\ $}}

\newcommand{\skipblk}[1]{}                                                      
\newcommand{\beqa}{\begin{eqnarray}}
\newcommand{\eeqa}{\end{eqnarray}} 
\newcommand{\bqa}{\begin{eqnarray}}
\newcommand{\eqa}{\end{eqnarray}}

\newcommand{\PR}[3]{{\em Phys. Rev.} {\bf #1}, #2 (19#3)}                             
\newcommand{\PL}[3]{{\em Phys. Lett.} {\bf #1}, #2 (19#3)}                            
\newcommand{\NP}[3]{{\em Nucl. Phys.} {\bf #1}, #2 (19#3)}                            
\newcommand{\PRL}[3]{{\em Phys. Rev. Lett.} {\bf #1}, #2 (19#3)}

\newcommand{\etal}{{\em et al., }}

\newcommand{\beq}{\begin{equation}}                                             
\newcommand{\eeq}{\end{equation}}

\newcommand{\RA}{\mbox{$\rightarrow$}}

\def\mxth{\mathsurround=0pt }
\def\xversim#1#2{\lower2.pt\vbox{\baselineskip0pt \lineskip-.5pt
  \ialign{$\mxth#1\hfil##\hfil$\crcr#2\crcr\sim\crcr}}}

\begin{document}

%\title{THEORETICAL SUMMARY, ELECTROWEAK PHYSICS}
\title{THEORETICAL SUMMARY, ELECTROWEAK PHYSICS\footnote{Talk presented
at the {\it $17^{th}$ International Workshop on Weak Interactions and
Neutrinos (WIN 99)}, Cape Town, South Africa, January 24-30, 1999.}}

\author{PAUL LANGACKER}

\address{Department of Physics and Astronomy \\ 
          University of Pennsylvania, Philadelphia PA 19104-6396, USA\\E-mail: 
pgl@langacker.hep.upenn.edu} 
%%%%%%%%%%%%%%%%%%%%%%%%%%%%%%%%%%%%%%%%%%%%%%%%%%%%%%%%%%%%%%

\maketitle\abstracts{Aspects of theoretical electroweak physics
are summarized, including the status of electroweak radiative corrections,
the hadronic contribution to the running of $\alpha$, global
fits to precision data and their implication for testing the standard
model and constraining new physics, and electroweak baryogenesis.}

\section{The $Z$, the $W$, and the weak neutral current}
The $Z$, the $W$, and the weak neutral current have always been
the primary tests of the unification part of the standard electroweak model.
Following the discovery of the neutral current in 1973,
its effects in pure weak processes such as $\nu N$ and $\nu e$
scattering and in weak-electromagnetic interference (e.g., $e D$ asymmetries,
$e^+ e^-$ annihilation, atomic parity violation) were intensely studied
in a series of experiments that were typically of several \% precision~\cite{history}.
The $W$ and $Z$ were discovered directly at CERN in 1983 and their
masses determined. In the 90's, the $Z$ pole experiments at LEP and the
SLC have allowed precision studies at the 0.1\% level of $M_Z$ (0.002\%) and
the $Z$ lineshape, branching ratios, and asymmetries; and recent measurements
at LEP II and the Tevatron have yielded $M_W$ to better than 0.1\%~\cite{nodulman}.

The implications of these results are
\begin{itemize}
\item The standard model is correct and unique to zeroth approximation,
confirming the gauge principle and the standard model gauge group
and representations.
\item The standard model is correct at the loop level, verifying 
the concept of renormalizable gauge theories, and allowing predictions
 from observed loop effects of $m_t$, $\alpha_s$, and $M_H$.
\item Possible new physics at the TeV scale is severely constrained,
strongly supporting such new physics as supersymmetry and unification, as opposed
to TeV-scale compositeness.
\item The gauge couplings at the electroweak scale are precisely determined,
allowing tests of gauge unification.
\end{itemize}

\section{Electroweak radiative corrections and
 the hadronic contribution to
$\alpha(M_Z)$} 
\label{theory}
J. Erler reviewed the status of electroweak radiation
corrections~\cite{erler,ew6}. Because of the accuracy of the high precision data,
multi-loop perturbative  calculations have to be performed. These include leading
two-loop electroweak, three-loop mixed electroweak-QCD, and three-loop QCD
corrections. 
${\cal O} (\alpha \alpha_s)$ vertex corrections to $Z$ 
decays~\cite{Czarnecki96A} have become available only recently, inducing 
an increase in the extracted $\alpha_s$ by about 0.001. The inclusion of top 
mass enhanced two-loop ${\cal O} (\alpha^2 m_t^4)$~\cite{Barbieri92} and 
${\cal O} (\alpha^2 m_t^2)$~\cite{Degrassi96} effects is crucial for a reliable
extraction of $M_H$. The latter, for example, lowers the extracted value of the
higgs mass by $\sim$ 18 MeV.

Erler has collected all available results in a new radiative correction
package. All $Z$ pole and low energy observables are self-consistently
evaluated with common inputs. The routines are written entirely within 
the \msb scheme, using \msb definitions for all gauge couplings and 
quark masses. This reduces the size of higher order terms in the QCD expansion.

The largest remaining theoretical uncertainty arises from the
$M_W$--$M_Z$--$\hat{s}^2_Z$ interdependence, where $\hat{s}^2_Z$ is
the weak angle in the \msb scheme. The problem is
directly related to the renormalization group  
running of the electromagnetic
coupling,
\beq
   \alpha (M_Z) = {\alpha\over 1 - \Delta\alpha (M_Z)}.
\eeq
While the contributions from leptons and bosons (and the top quark when not 
technically decoupled) can be computed with sufficient accuracy, the hadronic 
contributions from the five lighter quarks escape a first principle treatment 
due to strong interaction effects. M. Steinhauser~\cite{steinhauser}
reviewed the recent developments in the determination of $\alpha (M_Z)$
and the closely related problem of the hadronic contributions to the
anomalous magnetic moment of the muon~\cite{rn}.
These are calculated via dispersion relations involving the
cross section for $e^+ e^- \RA hadrons$. Early estimates  used
experimental data for the cross section up to around $\sqrt{s} \sim$ 40 GeV, and
perturbative QCD (PQCD) for higher energies. However, several groups
have emphasized that perturbative and nonperturbative QCD (using
sum rules and operator product expansions) are more reliable than the data
down to around 2 GeV, leading to a shifted value and smaller uncertainty.
Steinhauser described the impact of recent improved low energy data (e.g., 
below 1 GeV), as well the theoretical developments involving PQCD,
the charm threshold, QCD sum rules, and unsubtracted dispersion relations.
The recent calculations are in excellent agreement with each other,
and considerably reduce the theoretical uncertainties.

\section{Global fits and their implications}
J. Erler~\cite{erler} described the results of global fits
to all precision electroweak data, for testing the standard model,
determining its parameters, and searching for or constraining the
effects of new physics.
We used the complete data sets described in~\cite{erler,ew6}, and
carefully took into
account experimental and theoretical correlations, in particular in the $Z$-lineshape
sector, the heavy flavor  sector from LEP and the SLC, and for the deep
inelastic scattering experiments. 
Predictions within and beyond the SM were calculated by means of a new 
radiative correction program based on the $\msb$ renormalization scheme 
(see Section~\ref{theory}). All input and fit parameters are included in a 
self-consistent way, and the correlation (present in theory evaluations of 
$\alpha (M_Z)$) between $\alpha_s$ and the hadronic contribution is automatically
taken care of~\cite{je1}. We find very good agreement with the results of the 
LEPEWWG~\cite{Karlen98}, except for well-understood effects originating from
higher orders. We would like to stress that this agreement is quite remarkable
as they use the electroweak library ZFITTER~\cite{Bardin92}, which is based on 
the on-shell renormalization scheme. It also demonstrates that once the most 
recent theoretical calculations, in particular 
Refs.~\cite{Czarnecki96A,Degrassi96}, are taken into account, the theoretical 
uncertainty becomes quite small, and is in fact presently negligible compared 
to the experimental errors. The relatively large theoretical uncertainties 
obtained in the Electroweak Working Group Report~\cite{Bardin97} were 
estimated using different electroweak libraries, which did not include the 
full range of higher order contributions available now. 

In the Standard Model analysis we use the fine structure constant, $\alpha$,
and the Fermi constant, $G_F =  1.16637 (1) \times 10^{-5}$ GeV$^{-2}$, as 
fixed inputs. The error in $G_F$ is now of purely experimental origin after 
the very recent calculation of the two-loop QED corrections to $\mu$ decay have
been completed~\cite{vanRitbergen98}. They lower the central value by 
$2\times 10^{-10}$ GeV$^{-2}$ and the extracted $M_H$ by 1.3\%. Moreover, there
are five independent fit parameters, which can be chosen to be $M_Z$, $M_H$, 
$m_t$, $\alpha_s$, and the hadronic contribution to $\Delta\alpha (M_Z)$. 
Alternatively,
$M_Z$ can be  replaced by $s^2_W$ (the weak angle in the on-shell scheme) or
the \msb angle
$\hat{s}^2_Z$. We do not use 
$\alpha_s$ determinations from outside the $Z$ lineshape sector. The fit to all
precision data is perfect with an overall $\chi^2 = 28.8$ for 36 degrees of 
freedom, and yields~\cite{ew6},
\beq
\begin{array}{lcc}
\label{fitresults}
            M_H &=& 107^{+67}_{-45} \mbox{ GeV}, \\
           m_t &=& 171.4  \pm 4.8  \mbox{ GeV}, \\
      \alpha_s &=& 0.1206 \pm 0.0030, \\
   \hat{s}^2_Z &=& 0.23129 \pm 0.00019, \\
\bar{s}^2_\ell &=& 0.23158 \pm 0.00019, \\
         s^2_W &=& 0.22332 \pm 0.00045,
\end{array}
\eeq
where $\bar{s}^2_\ell \sim  \hat{s}^2_Z + 0.00029$ is the effective
angle usually quoted by the experimental groups~\cite{pdg}.
The larger uncertainty in the on-shell quantity $s^2_W$
is due to its greater sensitivity to $m_t$ and $M_H$.
None of the observables deviates from the SM best 
fit prediction by more than 2 standard deviations. 

The low value of of $M_H$ is consistent with the expectations
of supersymmetric extensions of the standard model in the
decoupling limit (for which the contributions of sparticles to
the radiative corrections are negligible). For a detailed discussion
of the upper limits on $M_H$ and their significance, see~\cite{erler}.
The value of $\alpha_s$ from the precision measurements is
consistent with other determinations~\cite{erler,ew6}. The precise
determination of $\hat{s}^2_Z$ and $\alpha_s$ allows a test
of gauge unification. The values are compatible with minimal
supersymmetric grand unified theories, when threshold corrections
at the high and low scales are included~\cite{GUT},
but not with the simplest
non-supersymmetric grand unified theories.

The precision data also allow stringent constraints on physics beyond the
standard model. Typically, one expects that new physics at the TeV scale that does
not decouple (i.e., the radiative corrections do not become smaller for larger
scales for the new physics) should lead to deviations at the few \% level, 
to be compared with the 0.1\% observations. This class includes most
versions of composite fermions and dynamical symmetry breaking. 
On the other hand physics which decouples, such as softly broken 
supersymmetry for sparticle masses $\gg M_Z$, are compatible with
the observations.
Specific constraints on heavy $Z'$ bosons and on supersymmetry,
and constraints on general parametrizations of classes of extensions
of the standard model (such as extended technicolor or higher-dimensional
Higgs representions), are extremely stringent, and are described in~\cite{erler}.

As one example, consider the $\rho$-parameter, defined 
by
\beq
  \rho_0 = {M_W^2 \over M_Z^2 \hat{c}^2_Z \hat{\rho} (m_t,M_H)},
\eeq
where $\hat{c}^2_Z \equiv 1 - \hat{s}^2_Z$, and $\hat{\rho}$
incorporates standard model radiative corrections. $\rho_0$
 is a measure of the neutral to charged current interaction strength. 
The SM contributions are absorbed in $\hat{\rho}$, so that in the SM 
$\rho_0 = 1$, by definition. Examples for sources of $\rho_0 \neq 1$ 
include non-degenerate extra fermion or boson doublets, and non-standard Higgs
representations. 

In a fit to all data with $\rho_0$ as an extra fit parameter, we obtain,
\beq
\begin{array}{lcr}
\label{rhofit}
         \rho_0 &=& 0.9996^{+0.0009}_{-0.0006}, \vspace{2pt} \\
            m_t &=& 172.9   \pm 4.8 \mbox{ GeV}, \\
       \alpha_s &=& 0.1212  \pm 0.0031,
\end{array}
\eeq
in excellent agreement with the SM. The central values are for $M_H = M_Z$, 
and the uncertainties are $1 \sigma$ errors and include the range,
$M_Z \leq M_H \leq 167$~GeV, in which the minimum $\chi^2$ varies within one 
unit. Note, that the uncertainties for $\ln M_H$ and $\rho_0$ are non-Gaussian:
at the $2 \sigma$ level ($\Delta \chi^2 \leq 4$), Higgs masses up to 800~GeV 
are allowed, and we find
\beq
   \rho_0 = 0.9996^{+0.0031}_{-0.0013} \mbox{ ($2 \sigma$)}.
\eeq
This implies strong constraints on the mass splittings of extra fermion and 
boson doublets~\cite{Veltman77}, 
\beq
  \Delta m^2 = m_1^2 + m_2^2 - \frac{4 m_1^2 m_2^2}{m_1^2 - m_2^2} 
               \ln {m_1\over m_2} \geq (m_1 - m_2)^2,
\eeq
namely, at the $1\sigma$ and $2\sigma$ levels, respectively,
\beq
\label{splittings}
   \sum\limits_i {C_i\over 3} \Delta m^2_i < \mbox{ (38 GeV)}^2 
   \mbox{ and (93 GeV)}^2,
\eeq
where $C_i$ is the color factor. 
Generalizations to the $S$, $T$, and $U$ parameters, which can describe
the effects of degenerate chiral fermions, are described in~\cite{erler,ew6,pdg}.

\section{Electroweak baryogenesis}
J. R. Espinosa surveyed the status of electroweak baryogenesis in the
standard model and its supersymmetric extension~\cite{espinosa}.
As is well known, a baryon asymmetry can be created cosmologically if the three
Sakharov conditions are satisfied: (1) baryon number violation; (2)
$C$ and $CP$ violation (to distinguish baryons from antibaryons); and (3)
thermal non-equilibrium in the baryon number violating processes.
Baryon number violation (with $B-L$ conserved) 
is present in the standard model as a
non-perturbative tunneling between degenerate vacua. The tunneling rate
is negligibly small at low temperature, but is enhanced by thermal
fluctuations for higher temperatures (``sphalerons''), 
especially above the electroweak
phase transition, for which the barrier height vanishes. Such effects
at and before the electroweak phase transition
would wash out any baryon asymmetry created at an earlier GUT era
if the latter have $B-L=0$. On the other hand, it is possible that a $B$ 
asymmetry was actually  created at the time of the electroweak transition,
as first discussed by Kuzmin, Rubakov, and Shaposhnikov~\cite{krs}.

The basic scenario is that if the electroweak transition is first order,
it proceeds by the creation and expansion of bubbles, with a
broken phase inside and an unbroken phase outside. Baryon number
violation can occur outside  the expanding bubble, where it is unsuppressed.
The $C$ and $CP$ breaking is manifested by $CP$-asymmetric reflection and transition
rates for massless fermions and antifermions as they encounter the expanding wall,
leading for example, to an excess of baryons entering the expanding bubble.
Necessary conditions for this to occur are not only sufficient $CP$ violation, but
also a first order transition, and finally that $v/T_c\ge 1$, where $v$
and $T_c$ are respectively the electroweak scale and the critical temperature
for the transition. If the latter is not satisfied, $B$ violation inside
the bubble will occur, destroying the asymmetry.

Espinosa surveyed the current situation. Within the standard model,
the conditions of a first order transition and $v/T_c\ge 1$
require respectively that the Higgs mass satisfies $M_H < 72$ GeV and
50 GeV, in contradiction with the experimental lower limit of around 97 GeV.

The situation is modified in the MSSM due to (1) new sources of $CP$
violation, (2) an extended Higgs sector with two doublets, and (3)
the influence of stops. Many authors~\cite{espinosa} have explored
the possibilities in detail. The upshot is that baryogenesis in the
MSSM is not excluded, but only works for a limited region of
parameter space which is explorable at LEP II and the Tevatron.
For example, the $v/T_c\ge 1$ condition requires $M_H < 105-110$ GeV,
$m_{\tilde{t}_R} < m_t$, small $\tan \beta$, and large $m_A$.

If these conditions are not satisfied, then baryogenesis would require new
mechanisms, such as extensions of the MSSM involving additional Higgs fields
or additional gauge symmetries. Another possibility, related to non-zero neutrino
mass, is that a  lepton asymmetry was created at an early epoch, e.g., 
by out of equiibrium decays of heavy Majorana neutrinos, and then converted to
a baryon asymmetry by the $B-L$ conserving sphaleron effects~\cite{lepton}.

\section*{Acknowledgments} 
This work was supported by U.S. Department of Energy Grant No. DOE-EY-76-02-3071.
It is a pleasure to thank the participants in the electroweak working 
group, and Jens Erler for collaboration.

\end{document}